\documentclass[times,tighten,twocolumn]{aastex631}

\usepackage{multirow}
\usepackage{graphicx}
\usepackage{subfigure}
\usepackage{color}
\usepackage{amsfonts}
\usepackage{mathrsfs}
\usepackage{amssymb}
\usepackage{hyperref}
\usepackage{ulem}
\usepackage{rotating} 

\usepackage{paralist}
\usepackage{amsmath}

\begin{document}

\title{Changing-look Active Galactic Nuclei from the Dark Energy Spectroscopic Instrument. IV. \\ Broad Emission Line Evolution Sequence Among H$\alpha$, Mg\,\textsc{ii}, and H$\beta$
}
    
    \author[0000-0001-9457-0589]{Wei-Jian Guo}
    \affiliation{Key Laboratory of Optical Astronomy, National Astronomical Observatories, Chinese Academy of Sciences, Beijing 100012, China\\ Email:\href{mailto:guowj@bao.ac.cn}{guowj@bao.ac.cn}}
    
    \author[0000-0003-1251-532X]{Victoria A. Fawcett}
    \affiliation{SSchool of Mathematics, Statistics and Physics, Newcastle University, Newcastle upon Tyne, NE1 7RU, UK}

    \author[0000-0002-2949-2155]{Małgorzata Siudek}

    \affiliation{Instituto de Astrof\'{\i}sica de Canarias, V\'{\i}a L\'actea, 38205 La Laguna, Tenerife, Spain}
    \affiliation{Instituto de Astrof\'isica de Canarias (IAC); Departamento de Astrof\'isica, Universidad de La Laguna (ULL), 38200, La Laguna, Tenerife, Spain}

    \author[0000-0001-5841-9179]{Yan-Rong Li}
    \affiliation{Key Laboratory for Particle Astrophysics, Institute of High Energy Physics, Chinese Academy of Sciences, 19B Yuquan Road, \\
    Beijing 100049, People's Republic of China}

    \author[0000-0003-0202-0534]{Cheng Cheng}
    \affiliation{Chinese Academy of Sciences South America Center for Astronomy, National Astronomical Observatories, CAS, Beijing 100101, \\ People's Republic of China }

    \author[0000-0002-5854-7426]{Swayamtrupta Panda}\thanks{Gemini Science Fellow} 
    \affiliation{ International Gemini Observatory/NSF NOIRLab, Casilla 603, La Serena, Chile}

    \author[0000-0003-0230-6436]{Zhiwei Pan}
    \affiliation{Kavli Institute for Astronomy and Astrophysics at Peking University, PKU, 5 Yiheyuan Road, Haidian District, Beijing 100871, P.R. China}

    \author[0000-0002-1234-552X]{shengxiu Sun}
    \affiliation{Kavli Institute for Astronomy and Astrophysics at Peking University, PKU, 5 Yiheyuan Road, Haidian District, Beijing 100871, P.R. China}

    \author[0000-0002-7719-5809]{Claire L. Greenwell}
    \affiliation{Institute for Computational Cosmology, Department of Physics, Durham University, South Road, Durham DH1 3LE, UK}

    \author[0000-0002-5896-6313]{David M. Alexander}

    \affiliation{Institute for Computational Cosmology, Department of Physics, Durham University, South Road, Durham DH1 3LE, UK}
    \affiliation{Centre for Extragalactic Astronomy, Department of Physics, Durham University, South Road, Durham, DH1 3LE, UK}

    \author[0000-0002-2733-4559]{John Moustakas}
    \affiliation{Department of Physics and Astronomy, Siena College, 515 Loudon Road, Loudonville, NY 12211, USA}

    \author[0009-0005-4152-2088]{Shuo Zhai}
    \affiliation{Key Laboratory of Optical Astronomy, National Astronomical Observatories, Chinese Academy of Sciences, Beijing 100012, China\\ Email:\href{mailto:guowj@bao.ac.cn}{guowj@bao.ac.cn}}

    \author[0000-0002-8402-3722]{Jun-Jie Jin}
    \affiliation{Key Laboratory of Optical Astronomy, National Astronomical Observatories, Chinese Academy of Sciences, Beijing 100012, China\\ Email:\href{mailto:guowj@bao.ac.cn}{guowj@bao.ac.cn}}

    \author[0000-0003-4200-9954]{Huaqing Cheng}
    \affiliation{Key Laboratory of Optical Astronomy, National Astronomical Observatories, Chinese Academy of Sciences, Beijing 100012, China\\ Email:\href{mailto:guowj@bao.ac.cn}{guowj@bao.ac.cn}}

    \author[0000-0002-0779-1947]{Jingwei Hu}
    \affiliation{Key Laboratory of Optical Astronomy, National Astronomical Observatories, Chinese Academy of Sciences, Beijing 100012, China\\ Email:\href{mailto:guowj@bao.ac.cn}{guowj@bao.ac.cn}}

    \author[0000-0003-4280-7673]{Yong-Jie Chen}
    \affiliation{Key Laboratory for Particle Astrophysics, Institute of High Energy Physics, Chinese Academy of Sciences, 19B Yuquan Road, \\
    Beijing 100049, People's Republic of China}

    \author[0000-0002-2419-6875]{Zhi-Xiang Zhang}
    \affiliation{College of Physics and Information Engineering, Quanzhou Normal University, Quanzhou, Fujian 362000, People’s Republic of China}

    \author[0000-0001-9449-9268]{Jian-Min Wang}
    \affiliation{Key Laboratory for Particle Astrophysics, Institute of High Energy Physics, Chinese Academy of Sciences, 19B Yuquan Road, \\
    Beijing 100049, People's Republic of China}
    \affiliation{School of Astronomy and Space Science, University of Chinese Academy of Sciences, 19A Yuquan Road, Beijing 100049, China}
    \affiliation{National Astronomical Observatories of China, Chinese Academy of Sciences, 20A Datun Road, Beijing 100020, China}

\received{xxx}
\revised{xxx}
\accepted{xxx}

\begin{abstract}

From a parent catalog of 561 changing-look active galactic nuclei (CL-AGNs) identified by \citet{Guo2025}, we investigate the evolutionary sequence of broad-emission lines using a redshift-selected subset ($0.35 < z < 0.45$)  of 54 CL-AGNs whose Dark Energy Spectroscopic Instrument (DESI) spectra simultaneously cover the H$\alpha$, H$\beta$, and Mg\,\textsc{ii} emission lines. To provide a baseline for comparison, we construct a control sample of 19,897 normal Type~1 AGNs within the same redshift range from the DESI Year~1 data. Through stacked spectral analysis and line–continuum luminosity correlations, we identify a clear evolutionary sequence in all AGN where broad H$\beta$ fades first, followed by Mg\,\textsc{ii}, and then H$\alpha$, as the AGN luminosity declines—consistent with expectations from reverberation mapping. This trend reflects a radially stratified broad line region (BLR), where each line’s responsivity depends on its ionization potential and radial distance from the central engine. In addition, we find that more massive supermassive black holes (SMBHs) require lower Eddington ratios to fully suppress broad emission lines, suggesting that the critical accretion threshold for the CL phenomenon is mass-dependent. Our results present the first statistical confirmation of a stratified broad line fading sequence in AGNs, reinforcing the central role of accretion state in shaping BLR structure and visibility.

\end{abstract}
 
\section{Introduction}

Active galactic nuclei (AGNs) exhibit a wide diversity of observational properties, which can largely be unified under a framework in which these differences arise from either orientation effects or changes in the accretion state of circumnuclear material \citep{Antonucci1993, Ho2008}. Among the most striking examples of AGN variability are the so-called ``changing-look'' AGNs (CL-AGNs), which transition between Type~1 (with broad emission lines) and Type~2 (with only narrow lines) spectral classifications---and vice versa---as broad-emission lines emerge or disappear \citep[e.g.,][]{Shappee2014, LaMassa2015, MacLeod2016, Yang2018}. These transitions are typically accompanied by substantial continuum variability from the optical to ultraviolet, as well as changes in multiple broad line species, including Balmer lines (e.g., H$\alpha$, H$\beta$) and low-ionization lines such as Mg\,\textsc{ii}, C\,\textsc{iii]}, and C\,\textsc{iv}. Because they trace rapid restructuring of the innermost regions of AGNs, CL-AGNs offer a unique opportunity to probe the physical conditions of the broad line region (BLR) and the nature of supermassive black holes (SMBHs) accretion.

Several physical mechanisms have been proposed to explain the CL phenomenon, including variable accretion rates, obscuration by intervening dust, and tidal disruption events \citep{Denney2014, Merloni2015, Ruan2016, Graham2020}. Among these, changes in accretion rate have gained the strongest observational and theoretical support. In particular, the disappearance of the soft X-ray excess---a key source of ionizing photons---in CL-AGNs \citep{Noda2018}, and the tight correlation between continuum dimming and weakening of broad-emission lines \citep{Green2022, Guo2025}, both point toward an accretion-driven origin. In this framework, a radiatively efficient, geometrically thin accretion disk is essential for sustaining the ionizing continuum that powers the BLR. Once the Eddington ratio drops below a critical threshold, structural changes in the disk suppress the ionizing photon output, leading to the disappearance of broad emission lines.

This picture is consistent with disk-wind models, in which the BLR is formed by radiatively or magnetically driven winds from the accretion disk \citep{Murray1995, Elitzur2014, Nomura2020}. In such models, the BLR is expected to vanish when the accretion rate falls below a critical value, due to insufficient radiation pressure to launch line-emitting gas. Theoretical arguments \citep{Nicastro2000, Elitzur2009} suggest that the BLR can only form above a minimum luminosity that depends on black hole mass. Photoionization simulations \citep{Baskin2018}, along with observational synthesis \citep{Stern2012}, further demonstrate that BLR properties---such as line width, ionization structure, and responsivity---are tightly coupled to the accretion flow.

While this framework explains the global disappearance of the BLR in CL-AGNs, it remains unclear whether individual broad-emission lines (e.g., H$\beta$, H$\alpha$, Mg\,\textsc{ii}) fade simultaneously or follow a sequential pattern in response to declining accretion power. Observational evidence from individual CL-AGN sources suggests that different lines may respond differently to continuum dimming \citep{Guo2024}, implying that each line traces distinct physical conditions within the BLR. Moreover, it remains an open question whether BLR evolution in CL-AGNs follows the same sequence as in normal AGNs undergoing more gradual variability. Understanding these differential line responses provides a unique diagnostic of the BLR’s radial stratification, ionization structure, and coupling to the central engine.

Photoionization models—particularly those based on the Locally Optimally Emitting Cloud (LOC) framework—predict that the BLR should ``collapse'' in an inside-out fashion as the ionizing continuum weakens \citep{Baldwin1995}. H$\beta$ are expected to disappear first due to their origin in compact, smaller ionized regions, followed by H$\alpha$, and then Mg\,\textsc{ii}, which arises from ionization gas at larger radii with lower responsivity \citep{guoh2020}. This theoretical sequence is supported by observational discoveries of CL-AGNs in which Mg\,\textsc{ii} fades only after Balmer lines have disappeared \citep{guohengxiao2019}, suggesting that line disappearance is governed jointly by ionization energy and BLR radius, modulated by the Eddington ratio.

However, previous studies on this topic have largely focused on individual case studies or specific photoionization simulations, limiting broader conclusions. A population-wide statistical test of the broad line fading sequence across a range of Eddington ratios and luminosities has remained elusive for both CL-AGNs and normal Type~1 AGNs. The advent of large-scale, homogeneous spectroscopic surveys such as the Dark Energy Spectroscopic Instrument (DESI) now provides the opportunity to conduct such a study \citep{DESI_Levi,DESI_2016_II,DESI_2022,DESI_SV}.

In this paper, we utilize the CL-AGN catalog constructed by \citet{Guo2025} from DESI Year~1 data, focusing on a redshift-selected subset ($0.35 < z < 0.45$) in which H$\alpha$, H$\beta$, and Mg\,\textsc{ii} are all simultaneously covered in the spectra. This enables, for the first time, a direct comparison of the fading behavior of these key broad lines across a wide range of accretion states in CL-AGNs. For reference, we construct a control sample of normal Type~1 AGNs in the same redshift range. Through stacked spectral analysis and luminosity correlation studies, we identify a clear evolutionary sequence in which broad H$\beta$ fades first, followed by Mg\,\textsc{ii}, and finally H$\alpha$.

This paper---the fourth in our DESI CL-AGN series---presents the first statistical confirmation of a stratified BLR fading sequence, interpreted in the context of BLR geometry, disk-wind models, and accretion physics. The remainder of this paper is organized as follows. In Section~\ref{sec:data}, we describe the DESI dataset, sample selection, and spectral measurements. Section~\ref{sec:results} and \ref{sec:discussion} presents our observational results and compares them with theoretical expectations. A summary is provided in Section~\ref{sec:conclusion}. Throughout this work, we adopt a flat $\Lambda$CDM cosmology with $H_0 = 67\,\mathrm{km\,s^{-1}\,Mpc^{-1}}$, $\Omega_\Lambda = 0.68$, and $\Omega_m = 0.32$, consistent with DESI Year~1 cosmological constraints \citep{DESI_2024_VI} and the \citet{Planck2020} results.

\begin{figure*}[ht]
    \centering
    \includegraphics[width=0.95\textwidth]{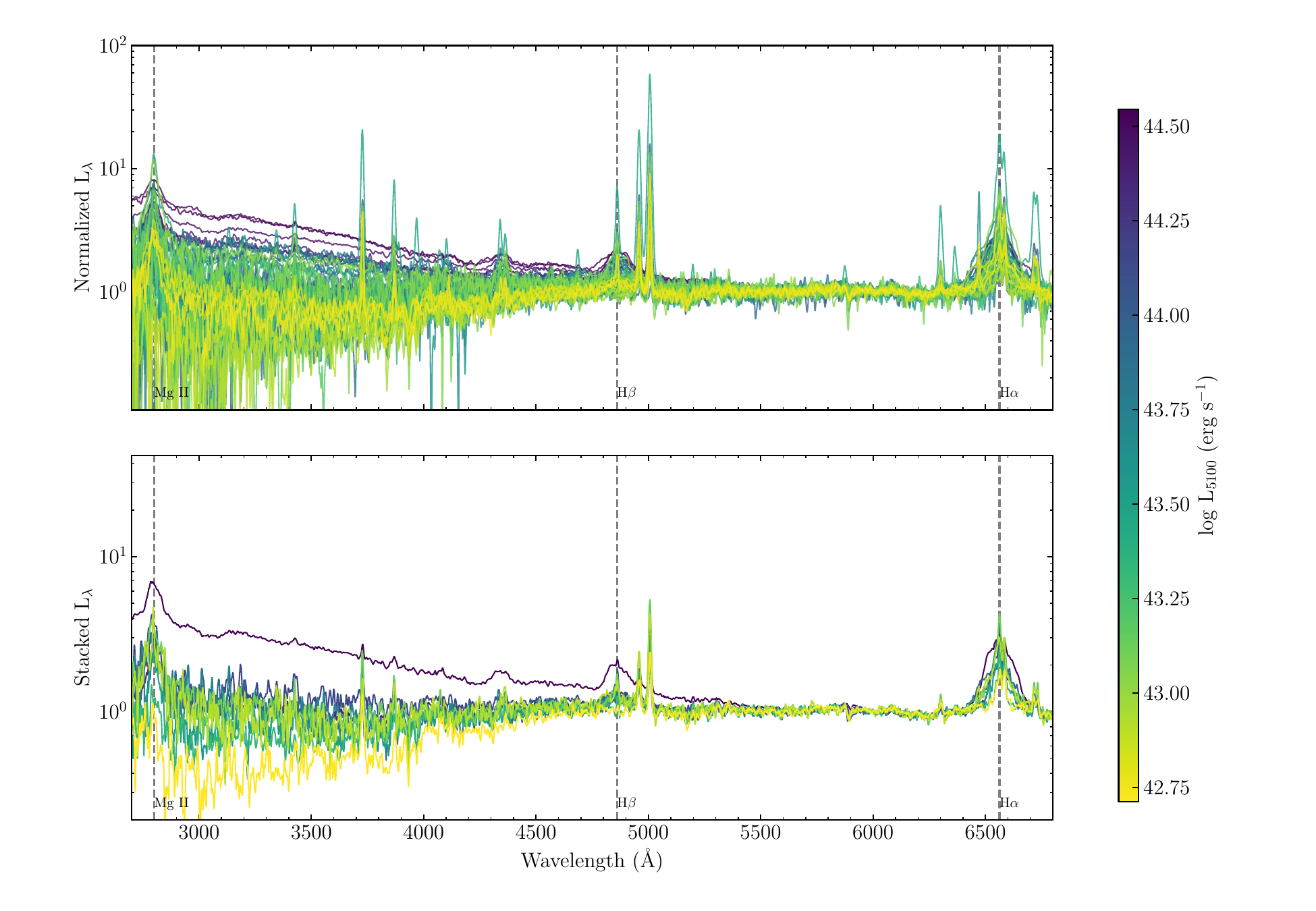}
    \caption{ Optical spectra of the CL-AGN core sample (top panel)  and  10 stacked spectra (bottom panel)  in the redshift range $0.35 < z < 0.45$, color-coded by continuum luminosity at 5100 \AA ($L_{5100}$). The top panel shows individual spectra normalized at 6000 \AA to highlight relative emission line strengths, while the bottom panel displays the corresponding stacked (mean-combined) spectra in luminosity bins. Prominent broad and narrow emission lines, including Mg\,\textsc{ii}, H$\beta$, and H$\alpha$, are visible.}
    \label{fig:stacked_spectra_CLAGN}
\end{figure*}

\begin{figure*}[ht]
    \centering
    \includegraphics[width=0.95\textwidth]{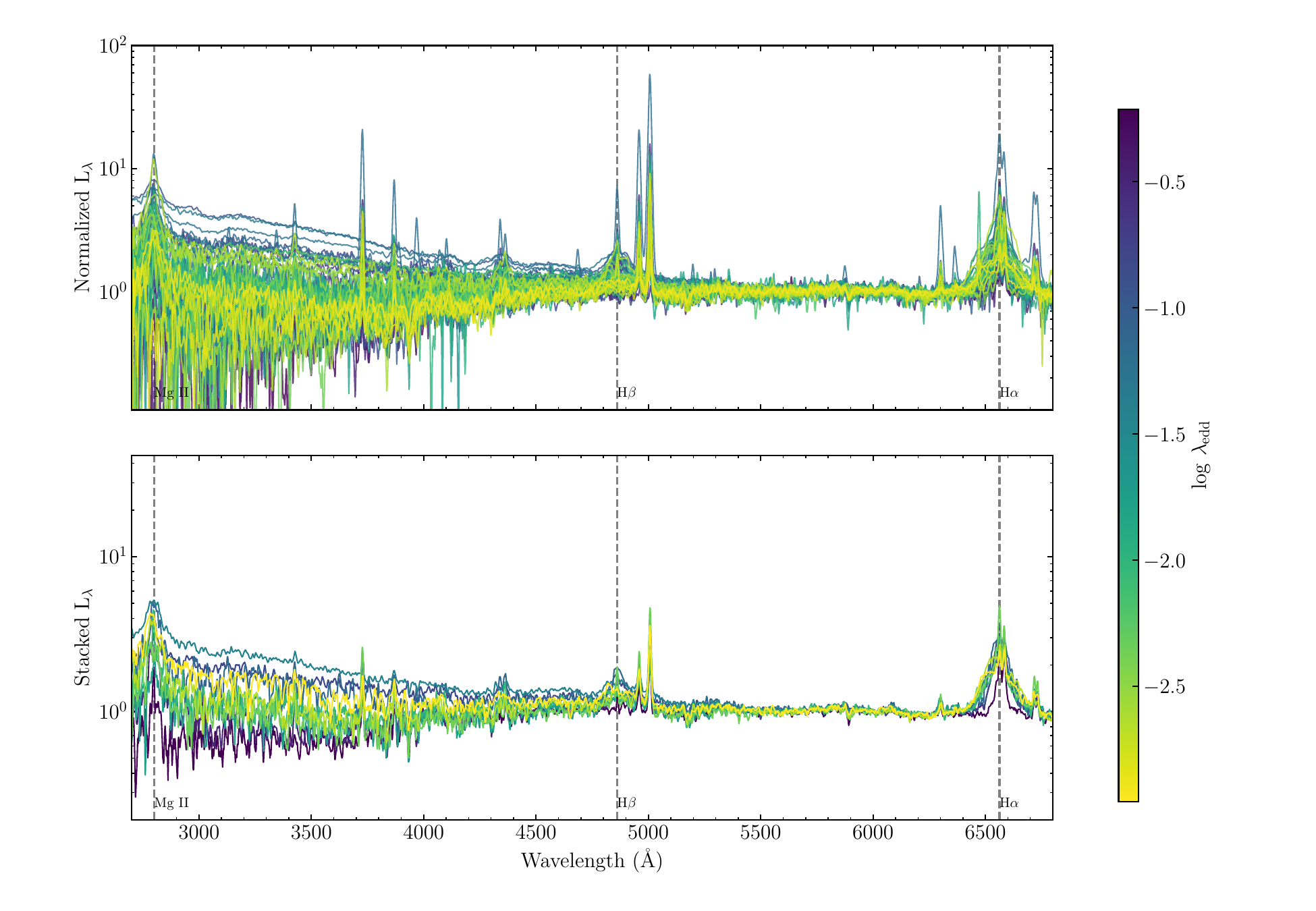}
    \caption{ Optical spectra of the CL-AGN core sample (top panel)  and  10 stacked spectra (bottom panel)  in the redshift range $0.35 < z < 0.45$, same with Figure \ref{fig:stacked_spectra_CLAGN} but color-coded by Eddington ratio   ($\lambda_{\rm edd}$).}
    \label{fig:stacked_spectra_CLAGN2}
\end{figure*}

\section{Sample and Measurement}
\label{sec:data}

\subsection{CL-AGN: Parent Sample and Core Sample }

The CL-AGN parent sample comprises 561 sources identified by \citet{Guo2025} from the first data release of DESI \citep{DESI_DR1}. These objects were selected by comparing DESI spectra with archival SDSS observations, requiring the appearance or disappearance of broad-emission lines (H$\beta$, H$\alpha$, or Mg\,\textsc{ii}), with careful consideration of flux calibration. The sample includes 527 cases with H$\beta$ variability, 149 with H$\alpha$, and 129 showing changes in Mg\,\textsc{ii}.

Due to DESI’s broad spectral coverage ($3600$–$9800$\,\AA), all three lines—H$\alpha$, H$\beta$, and Mg\,\textsc{ii}—are simultaneously observable in the redshift range $0.35 \leq z \leq 0.45$. We therefore define a core sample of 55 CL-AGNs  with DESI spectra (40 dim state and 15 bright state) in this range, enabling direct, line-by-line comparisons across all three broad lines. After excluding one object with incomplete spectral coverage, the final core sample consists of 54 CL-AGNs.

We rank the 54 CL-AGNs by their continuum luminosity at 5100\,\AA\ in descending order and display the individual, normalized spectra (scaled at 6000\,\AA) in the top panel of Figure~\ref{fig:stacked_spectra_CLAGN}. To highlight systematic trends in emission-line behavior, we further divide the sample into 10  luminosity ($L_{5100}$) bins  and stack $5\sim6$  spectra within each bin. The resulting composite spectra are shown in the bottom panel of Figure~\ref{fig:stacked_spectra_CLAGN}.

The stacked spectra exhibit clear luminosity-dependent trends: low-luminosity sources (yellow curves) show redder continua and weaker broad lines, while high-luminosity sources (dark purple) display bluer slopes and stronger emission features. This pattern reflects spectral hardening with increasing accretion power, consistent with a rising ionizing continuum at higher Eddington ratios.

Across the full luminosity range, the three broad-emission lines display distinct evolutionary behavior:
\begin{enumerate}
    \item H$\beta$ responds most sensitively to continuum luminosity, disappearing rapidly in low-$L_{5100}$ bins;
    \item H$\alpha$ remains detectable across a broader luminosity range, although it gradually weakens;
    \item Mg\,\textsc{ii} persists throughout all bins, showing a slower decline with relatively minor structural changes.
\end{enumerate}

To trace the accretion state more directly and to avoid mixing objects of similar luminosity but different black-hole masses, we additionally group the same 54 CL-AGNs by Eddington ratio, $\log\lambda_{\rm Edd}$. As in Figure~\ref{fig:stacked_spectra_CLAGN}, the objects are sorted and split into ten equal-occupancy bins (5–6 spectra per bin), using the same 6000\,\AA\ normalization and stacking procedure. The individual spectra are color-coded by $\log\lambda_{\rm Edd}$, and the corresponding stacks are shown in Figure~\ref{fig:stacked_spectra_CLAGN2}.

In the Eddington-ratio space, the three broad lines (Mg {\sc ii}, H$\beta$, and H$\alpha$), still fade monotonically with decreasing accretion rate. A noteworthy feature is that the lowest-luminosity stacks (purple curves in the bottom panel) nonetheless include sources with relatively high $\lambda_{\rm Edd}$, because these objects host comparatively small black holes—hence modest $L_{5100}$ does not necessarily imply a low accretion state.  To determine whether the sequence is governed primarily by luminosity or by Eddington ratio, we therefore perform a controlled comparison with a Type~1 AGN sample below.

\begin{figure*}[ht]
    \centering
    \includegraphics[width=\textwidth]{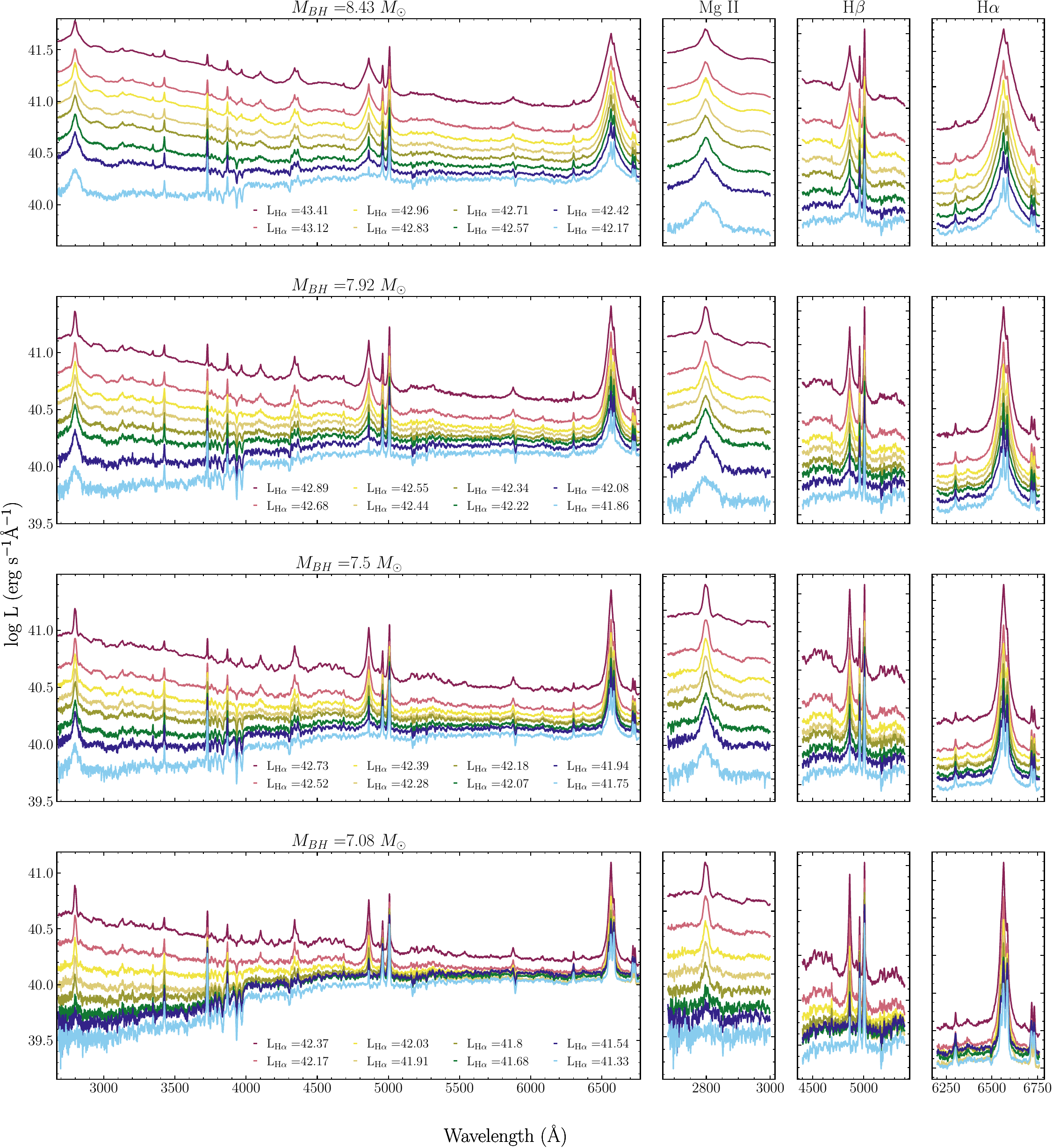}
    \caption{Composite spectra of control sample (Type~1 AGNs)  in four black hole mass ($M_{\rm BH}$) bins, further subdivided by  eight H$\alpha$ luminosity or continuum luminosity at 5100 \AA ($L_{5100}$). The leftmost panels show the full spectral range, while the right panels highlight the emission line regions of Mg\,\textsc{ii}, H$\beta$, and H$\alpha$. Each stacked curve corresponds to a median spectrum in a specific H$\alpha$ bin, as labeled. The color coding follows a gradient palette to ensure clear visual differentiation between luminosity bins.    
    }
    \label{fig:normalAGN_stacks}
\end{figure*}

\subsection{Control Sample: Type 1 AGNs}

To establish a baseline for comparison, we construct a control sample of normal Type~1 AGNs from the DESI DR1 dataset \citep{DESI_DR1}. Candidate quasars are selected using spectral classifications from the DESI pipeline \texttt{Redrock}, requiring \texttt{SPECTYPE = QSO} to ensure reliable identification. Additional quality cuts include \texttt{COADD\_FIBERSTATUS = 0} (indicating no known fiber issues) and \texttt{ZWARN} values of 0 or 4, the latter corresponding to minor redshift ambiguities that do not compromise line measurements.

We further restrict the sample to $0.35 < z < 0.45$ to ensure complete spectral coverage of H$\alpha$, H$\beta$, and Mg\,\textsc{ii}, yielding an initial set of 23,994 spectra. Spectral measurements are adopted from the \texttt{FastSpecFit} pipeline\footnote{\url{https://fastspecfit.readthedocs.io}}, a fast, automated fitting tool optimized for DESI data \citep{Moustakas2023}. From these outputs, we exclude sources where the fitted broad H$\alpha$ flux is consistent with zero, resulting in a control sample of 19,897 bona fide broad line AGNs. For this control sample, we extract the full width at half maximum (FWHM) and flux of the broad H$\alpha$ line to estimate black hole masses using virial scaling relations and to compute bolometric luminosities (see Section~\ref{sec:decomposition}).

To match the CL-AGN sample, we divide the control sample into four black hole mass bins ($\log M_{\rm BH}/{\rm M}_\odot = 7.08$, 7.50, 7.92, 8.43) and eight $\rm{H}\alpha$ luminosity ($L_{\mathrm{H}\alpha}$) bins, with $\sim$622 spectra per bin to ensure statistical uniformity. Median-stacked composite spectra are constructed for each bin.

As shown in Figure~\ref{fig:normalAGN_stacks}, both the continuum and broad emission-line strengths decrease systematically with decreasing $L_{\mathrm{H}\alpha}$ or $L_{5100}$, consistent with photoionization expectations. The stacked spectra reveal a coherent fading sequence:
\begin{itemize}
    \item H$\beta$ shows the fastest decline in both luminosity and width;
    \item H$\alpha$, while generally stronger, weakens gradually but remains detectable even in the faintest bins;
    \item Mg\,\textsc{ii} exhibits a slow but steady decline in peak intensity and equivalent width, with minimal change in profile shape.
\end{itemize}

\begin{figure*}
\centering
    \includegraphics[width=0.85\textwidth]{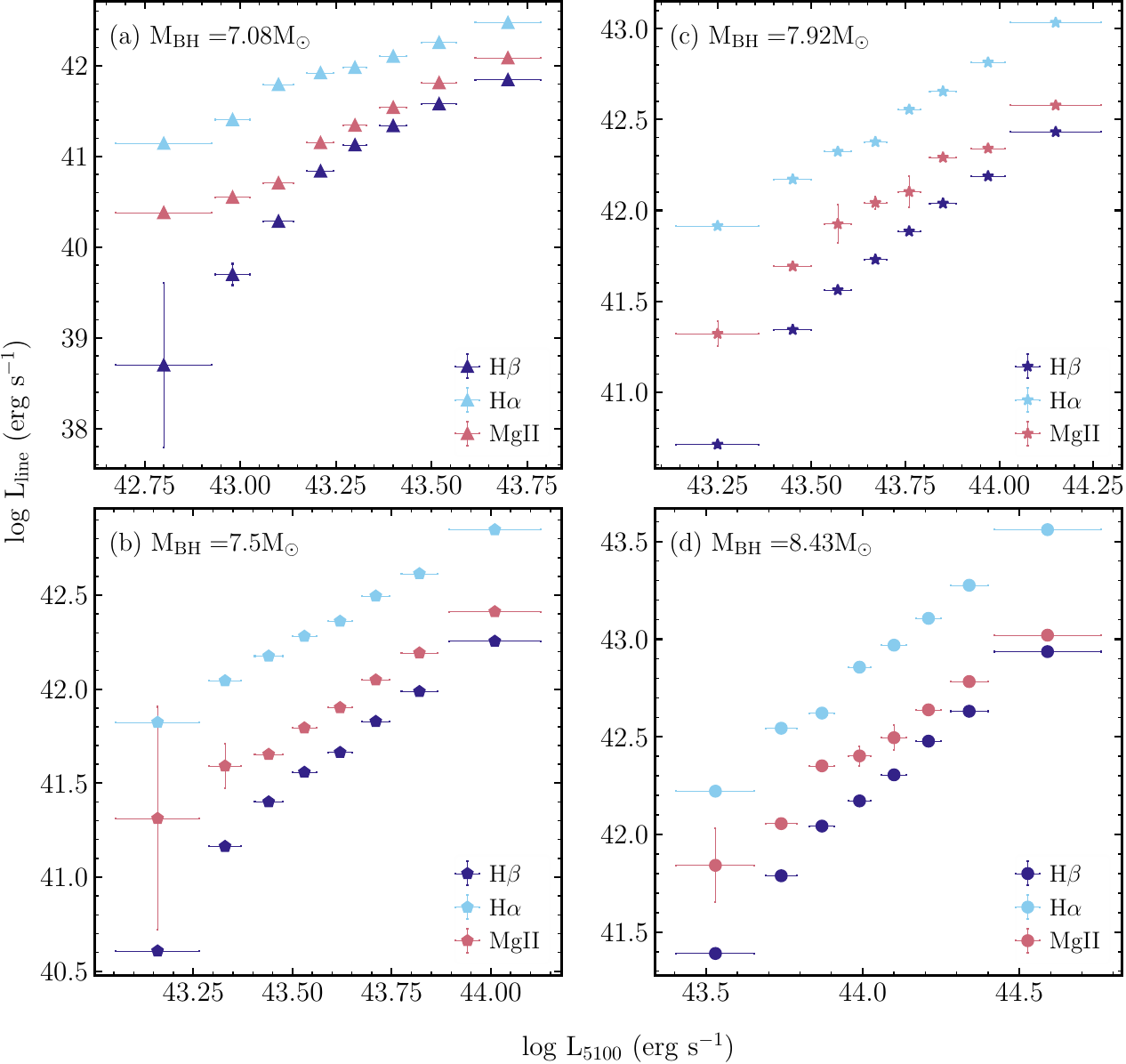}
\caption{
    The relation between the broad emission line luminosities and the continuum luminosity at 5100\,\AA\ ($L_{5100}$)  for four black hole mass bins. Each panel represents a different black hole mass: (a) \(\log (M_{\rm BH}/M_{\odot}) = 7.08\), (b) \(\log (M_{\rm BH}/M_{\odot}) = 7.50\), (c) \(\log (M_{\rm BH}/M_{\odot}) = 7.92\), and (d) \(\log (M_{\rm BH}/M_{\odot}) = 8.43\). The blue, cyan, and red points represent H$\beta$, H$\alpha$, and Mg\,\textsc{ii} emission lines, respectively. Error bars in both x- and y-directions denote the measurement uncertainties.}
\label{fig:line_luminosity2}
\end{figure*}

\subsection{Physical Property Measurements and Spectral Decomposition}
\label{sec:decomposition}
For the spectra of control sample (Type~1 AGN), we estimated continuum luminosities, black hole masses, and Eddington ratios based on broad H$\alpha$ measurements from the \texttt{FastSpecFit} pipeline. Specifically, we used the broad H$\alpha$ luminosity and FWHM ($V_{\rm H\alpha}$) to infer the monochromatic luminosity at 5100\,\AA\ ($L_{5100}$), which is not directly fitted by \texttt{FastSpecFit}. We adopted the empirical relation from \citet{Greene2005}:

\begin{align}
L_{\mathrm{H}\alpha} =\ & (5.25 \pm 0.02) \times 10^{42} \nonumber \\
& \times \left( \frac{L_{5100}}{10^{44}\ \mathrm{erg\ s}^{-1}} \right)^{1.157 \pm 0.005}\ \mathrm{erg\ s}^{-1}.
\end{align}

The broad line region (BLR) radius was then computed using the $R_{\rm H\alpha}$–$L_{5100}$ size–luminosity relation from reverberation mapping (RM) studies \citep{Cho2023}:

\begin{equation}
\log(R_{\rm H\alpha}/{\rm ld}) = K_2 + \alpha_2 \log \ell_{44},
\end{equation}

where $\ell_{44} = L_{5100}/10^{44}\,\mathrm{erg\ s^{-1}}$, and the best-fit coefficients are $K_2 = 1.59 \pm 0.05$ and $\alpha_2 = 0.58 \pm 0.04$.

The black hole mass was estimated using the standard single-epoch virial formula:

\begin{equation}
M_{\bullet} = f_{\rm BLR} \frac{V_{\rm BLR}^{2} R_{\rm BLR}}{G},
\label{eq_mass}
\end{equation}

where $f_{\rm BLR}$ is the virial scaling factor, $V_{\rm BLR}$ is the FWHM of the broad H$\alpha$ line, and $R_{\rm BLR}$ is the BLR radius derived above. We adopted $f_{\rm BLR} = 1.12$, following \citet{Woo2015}.

The continuum luminosity $L_{5100}$ was converted to a bolometric luminosity using a correction factor $C_{\rm bol} = 9.26$ \citep{Shen2008}. The Eddington ratio was then calculated as:

\begin{align}
\lambda_{\rm Edd} = \frac{L_{\rm bol}}{L_{\rm Edd}} = \frac{C_{\rm bol} \cdot L_{5100}}{L_{\rm Edd}}, \\ 
L_{\rm Edd} = 1.26 \times 10^{38} \left(\frac{M_{\rm BH}}{M_\odot}\right)\, \mathrm{erg\ s^{-1}}.
\end{align}

For each bin in black hole mass and H$\alpha$ luminosity, we adopted the median and 1$\sigma$ scatter of the derived parameters ($L_{\rm H\alpha}$, $L_{5100}$, $M_{\rm BH}$, and $\lambda_{\rm Edd}$) to represent the bin’s central value and uncertainty. This statistical approach effectively characterizes the parameter distributions while minimizing the influence of outliers. The results are consistent with those obtained from direct fits to the stacked spectra.

Since \texttt{FastSpecFit} does not model Fe\,\textsc{ii} emission, its measurements of H$\beta$ and Mg\,\textsc{ii} luminosities may be biased due to blending with Fe\,\textsc{ii} features. To improve accuracy, we performed detailed spectral decomposition using \texttt{DASpec} \citep{Du2024}, which explicitly models Fe\,\textsc{ii} and other components.

Following \cite{Guo2025}, we defined three fitting windows—2700--3000\,\AA, 3700--3800\,\AA, and 4300--6750\,\AA—corresponding to the Mg\,\textsc{ii}, H$\beta$, and H$\alpha$ regions, respectively, while avoiding the Balmer continuum. A 2.5\,Gyr stellar population template from \citet{Bruzual2003} was adopted to model the host galaxy contribution.

Each spectral fit included the following components:
\begin{itemize}
    \item A 2.5\,Gyr stellar population template to represent the host galaxy;
    \item A power-law continuum for the AGN;
    \item An Fe\,\textsc{ii} pseudo-continuum template from \citet{Boroson1992};
    \item Double-Gaussian models for the broad components of Mg\,\textsc{ii} $\lambda\lambda$2796,2803, H$\gamma$ $\lambda$4340, H$\beta$ $\lambda$4861, and H$\alpha$ $\lambda$6563;
    \item Single-Gaussian models for the narrow components of [O\,\textsc{ii}] $\lambda$3727, H$\gamma$ $\lambda$4340, H$\beta$ $\lambda$4861, [O\,\textsc{iii}] $\lambda\lambda$4959,5007, [O\,\textsc{i}] $\lambda$6300, [N\,\textsc{ii}] $\lambda\lambda$6548,6583, H$\alpha$ $\lambda$6563, and [S\,\textsc{ii}] $\lambda\lambda$6716,6731.
\end{itemize}

To reduce degeneracies and ensure consistency across spectral regions, we tied velocity parameters (redshift and FWHM) for the following groups:
\begin{itemize}
    \item The broad components of two Mg\,\textsc{ii} were tied together
    \item The broad H$\beta$ and H$\alpha$ were tied together;
    \item Narrow-line components were grouped as follows: \\
    (1) [O\,\textsc{ii}] $\lambda$3727 and H$\gamma$ $\lambda$4340; \\
    (2) H$\beta$ $\lambda$4861 and [O\,\textsc{iii}] $\lambda\lambda$4959,5007; \\
    (3) [O\,\textsc{i}] $\lambda$6300, [N\,\textsc{ii}] $\lambda\lambda$6548,6583, H$\alpha$ $\lambda$6563, and [S\,\textsc{ii}] $\lambda\lambda$6716,6731.
\end{itemize}

The results for the CL-AGN parent sample are adopted directly from \cite{Guo2025}. For the CL-AGN core sample, however, we apply the fitting approach described above with additional refinements from  \cite{Guo2025} for the  54 CL-AGNs in Figure \ref{fig:stacked_spectra_CLAGN}.

\begin{figure}
\centering
\includegraphics[width=0.47\textwidth]{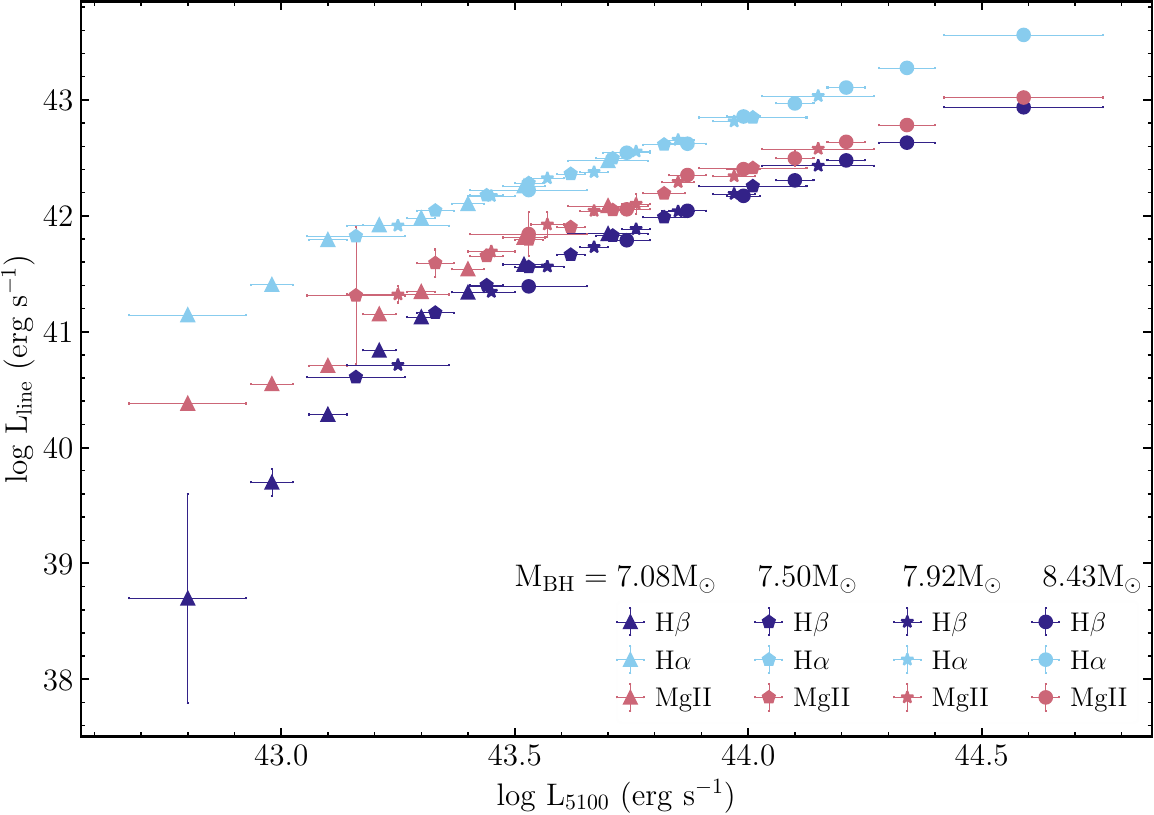}
\caption{ Correlation between the line luminosity ($L_{\rm line}$) and the continuum luminosity at 5100 \AA \  ($L_{5100}$) for quasars in four black hole mass ($M_{\rm BH}$) bins. Each panel represents one $M_{\rm BH}$ bin, and data points correspond to the median luminosities of H$\beta$ (dark blue circles), H$\alpha$ (light blue triangles), and Mg\,\textsc{ii} (red pentagons). Error bars denote the 16th–84th percentile range in each bin. }
\label{fig:line_luminosity}
\end{figure}

\begin{figure*}
\centering
\includegraphics[width=0.45\textwidth]{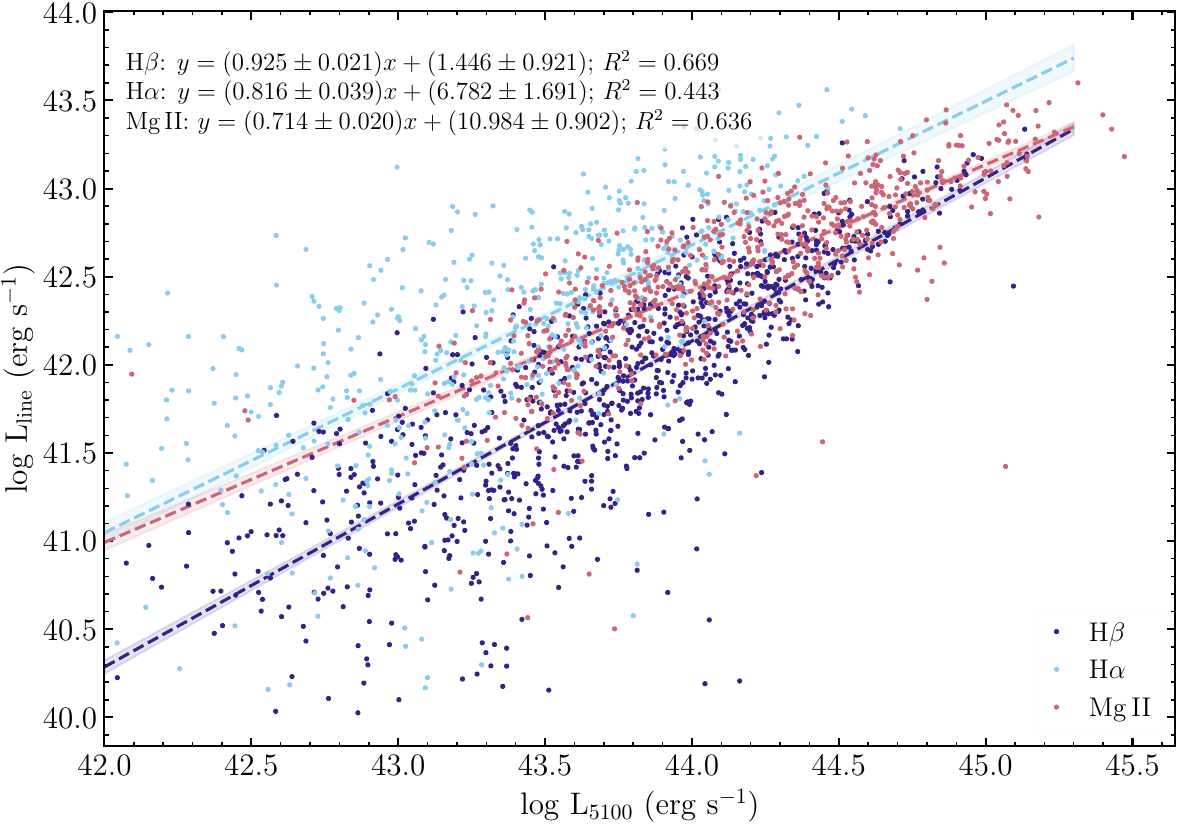}\hspace{0.5cm}
\includegraphics[width=0.45\textwidth]{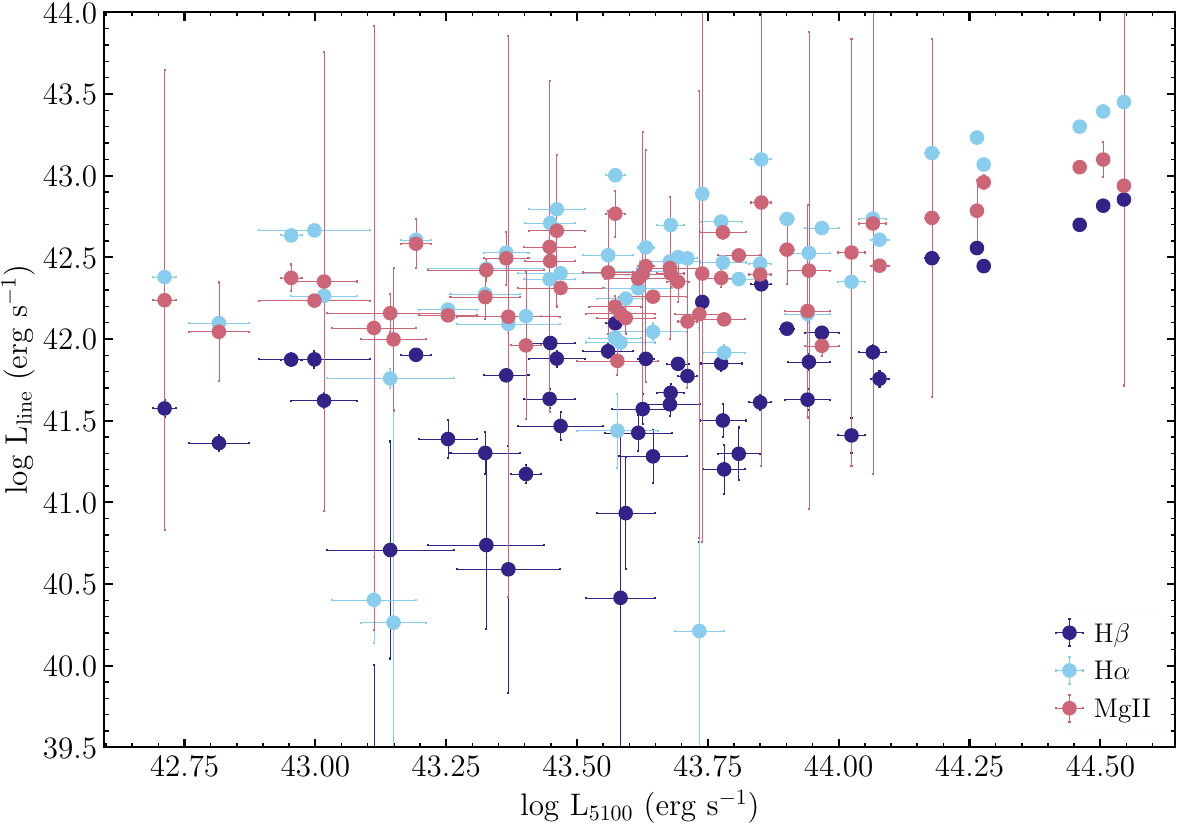}
\caption{
Correlation between the broad emission-line luminosity ($L_{\rm line}$) and the continuum luminosity for  CL-AGN  parent sample (left panel) and  CL-AGN core sample (right panel). Three broad lines—H$\beta$ (dark blue), H$\alpha$ (light blue), and Mg\,\textsc{ii} (red)— and their linear regressions are plotted for individual objects.\label{fig:line_luminosity_CLAGN}}
\end{figure*}

\section{Results}
\label{sec:results}

\subsection{Correlation Between Broad Emission Line Luminosities and Continuum Luminosities}

To investigate the relative response of H$\alpha$, H$\beta$, and Mg\,\textsc{ii} to continuum variations, we examine the correlations between broad emission-line luminosities and the continuum luminosity in both CL-AGNs and a control sample of normal Type~1 AGNs.

Figure~\ref{fig:line_luminosity2} and Figure~\ref{fig:line_luminosity} presents the $L_{\rm line}$–$L_{5100}$ relations for normal Type~1 AGNs, subdivided into four black hole mass ($M_{\rm BH}$) bins. Each panel displays the median $L_{\rm line}$ values across $L_{5100}$ bins at fixed $M_{\rm BH}$, providing a baseline against which CL-AGN behavior can be compared. All three lines exhibit strong positive correlations with $L_{5100}$, consistent with photoionization, and the trends appear largely independent of black hole mass.

However, the strength and slope of these correlations differ among the three lines. This difference reflects the well-known stratification of the BLR, which has also been established through RM studies. Specifically, RM-based radius–luminosity ($R$–$L$) relations indicate that the line response to continuum luminosity varies systematically by species. Recent measurements yield:
\begin{itemize}
    \item Mg\,\textsc{ii}:  $\log \tau_{\rm {rest}} = (0.31\pm0.06)\log { {L_{3000}}} + (2.086\pm0.03)$ \citep{Shen2024},
    \item H$\beta$: $\log \tau_{\rm {rest}} = (0.41\pm0.07)\log { {L_{5100}}} + (1.458\pm0.04)$ \citep{Shen2024},
    \item H$\alpha$: $\log \tau_{\rm {rest}} = (0.58\pm0.04)\log { {L_{5100}}} + (1.59\pm0.05)$ \citep{Cho2023},
\end{itemize}
where $\log \tau_{\rm rest}$ represents the rest-frame time lag for different lines, $L_{3000}$ denotes the monochromatic luminosity at 3000\,\AA, and $L_{5100}$ denotes the monochromatic luminosity at 5100\,\AA. Taken together, these relations imply that H$\alpha$ exhibits the strongest luminosity dependence (slope $\sim$0.58), followed by H$\beta$ (slope $\sim$0.41), and then Mg\,\textsc{ii} (slope $\sim$0.31). This ordering reflects a radially stratified BLR, where H$\alpha$ arises from a region whose effective size scales more steeply with luminosity than that of H$\beta$ or Mg\,\textsc{ii}. The trends observed in Figure~\ref{fig:line_luminosity2} and Figure~\ref{fig:line_luminosity}, including the steepest slope for H$\alpha$ and the flattest for Mg\,\textsc{ii}, are consistent with these expectations. The resulting decline sequence in broad line luminosities—H$\beta \rightarrow$ Mg\,\textsc{ii} $\rightarrow$ H$\alpha$—traces the fading of the ionizing continuum from the inside out.

Figures~\ref{fig:line_luminosity_CLAGN} extend this comparison to CL-AGNs using both  parent sample and core sample. CL-AGNs follow the same qualitative ordering as normal AGNs but exhibit systematic deviations at low continuum luminosities. In these faint states, H$\beta$ often drops most dramatically, frequently falling well below the baseline relation, whereas Mg\,\textsc{ii} remains relatively strong in some CL-AGNs. This behavior suggests that the BLR response in CL-AGNs broadly aligns with the $R$–$L$ expectations but is complicated by additional effects such as different ionization processes, anisotropic illumination, delayed response, or obscuration \citep{Ruan2016, Stern2018, Panda2021}. We discuss these outliers in more detail in Section~\ref{sec:outlier}.

To make the comparison explicit, we fit linear relations in log--log space, $\log L_{\rm line} = a + b\,\log L_{5100}$, to the points in Fig.~\ref{fig:line_luminosity_CLAGN}. Using ordinary least squares with $10^4$ bootstrap resamples for uncertainties, we obtain:
\[
\begin{aligned}
\mathrm{H}\beta &: \quad b = 0.92 \pm 0.02,\ \ a = 1.45 \pm 0.92,\\ & R^2=0.67,\ \ p<10^{-3},\\
\mathrm{H}\alpha &: \quad b = 0.82 \pm 0.04,\ \ a = 6.78 \pm 1.69,\\ & R^2=0.44,\ \ p<10^{-3},\\
\mathrm{Mg\,\textsc{ii}} &: \quad b = 0.71 \pm 0.02,\ \ a = 10.98 \pm 0.90,\\ & R^2=0.64,\ \ p<10^{-3}.
\end{aligned}
\]
The slope ordering is $b_{\mathrm{H}\beta} > b_{\mathrm{H}\alpha} > b_{\mathrm{Mg\,\textsc{ii}}}$, and the differences are statistically significant.
These fits (with 68\%/95\% confidence bands) are overplotted in Fig.~\ref{fig:line_luminosity_CLAGN}. Since $L_{\rm line}\propto L_{5100}^{\,b}$, the intrinsic Baldwin effect implies ${\rm EW}\propto L_{5100}^{\,b-1}$.
Our best-fit slopes therefore give $b-1=-0.075\pm0.021$ (H$\beta$), $-0.184\pm0.039$ (H$\alpha$), and $-0.286\pm0.020$ (Mg\,\textsc{ii}), indicating that Mg\,\textsc{ii} shows the strongest equivalent-width anti-correlation with continuum luminosity, while H$\beta$ responds most nearly proportionally to $L_{5100}$.
The comparatively low $R^2$ for H$\alpha$ suggests larger intrinsic dispersion, plausibly driven by optical-depth and collisional-excitation effects or extinction and host-continuum contamination that differentially affect $L_{5100}$ and Balmer lines, or non-simultaneity and light-travel-time delays that are especially relevant for CL-AGNs.

Our results are consistent with findings from RM and variability studies. It is well established that Mg\,\textsc{ii} exhibits weaker and slower variability than H$\beta$, reflecting its lower ionization potential and larger BLR radius \citep[e.g.,][]{Cackett2015, Sun2015}. Furthermore, the relatively weak “breathing” behavior of Mg\,\textsc{ii}, in contrast to the Balmer lines, has been observed in both low- and high-redshift quasars \citep[e.g.,][]{Homayouni2020}.

\begin{figure*}
\centering
    \includegraphics[width=0.85\textwidth]{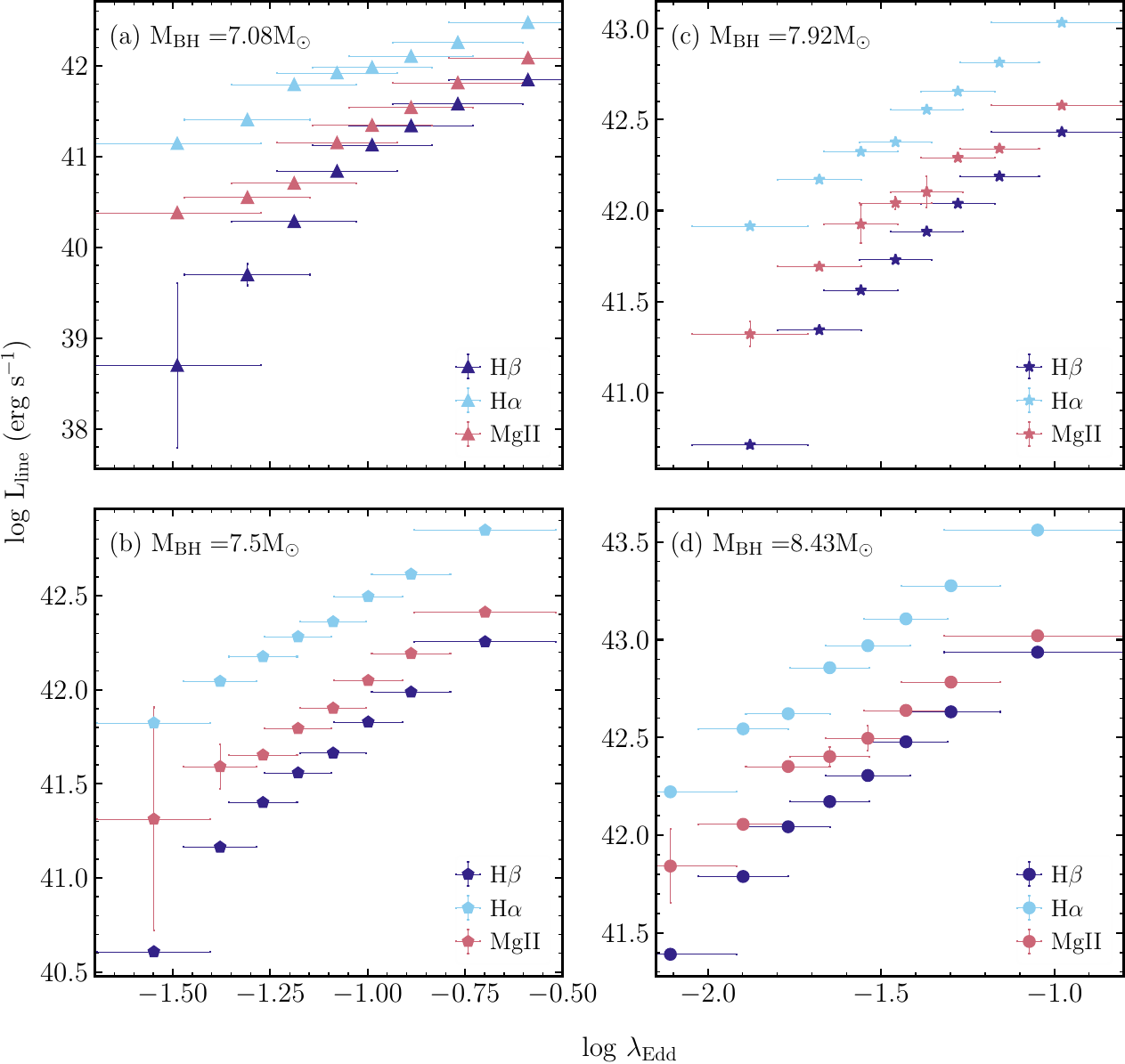}
\caption{
    The relation between the broad emission line luminosities and the Eddington ratios (\(\log \lambda_{\rm Edd}\)) for four black hole mass bins. Each panel represents a different black hole mass: (a) \(\log (M_{\rm BH}/M_{\odot}) = 7.08\), (b) \(\log (M_{\rm BH}/M_{\odot}) = 7.50\), (c) \(\log (M_{\rm BH}/M_{\odot}) = 7.92\), and (d) \(\log (M_{\rm BH}/M_{\odot}) = 8.43\). The blue, cyan, and red points represent H$\beta$, H$\alpha$, and Mg\,\textsc{ii} emission lines, respectively. Error bars in both x- and y-directions denote the measurement uncertainties.}
\label{fig:line_eddtion}
\end{figure*}

\begin{figure*}
\centering
\includegraphics[width=0.85\textwidth]{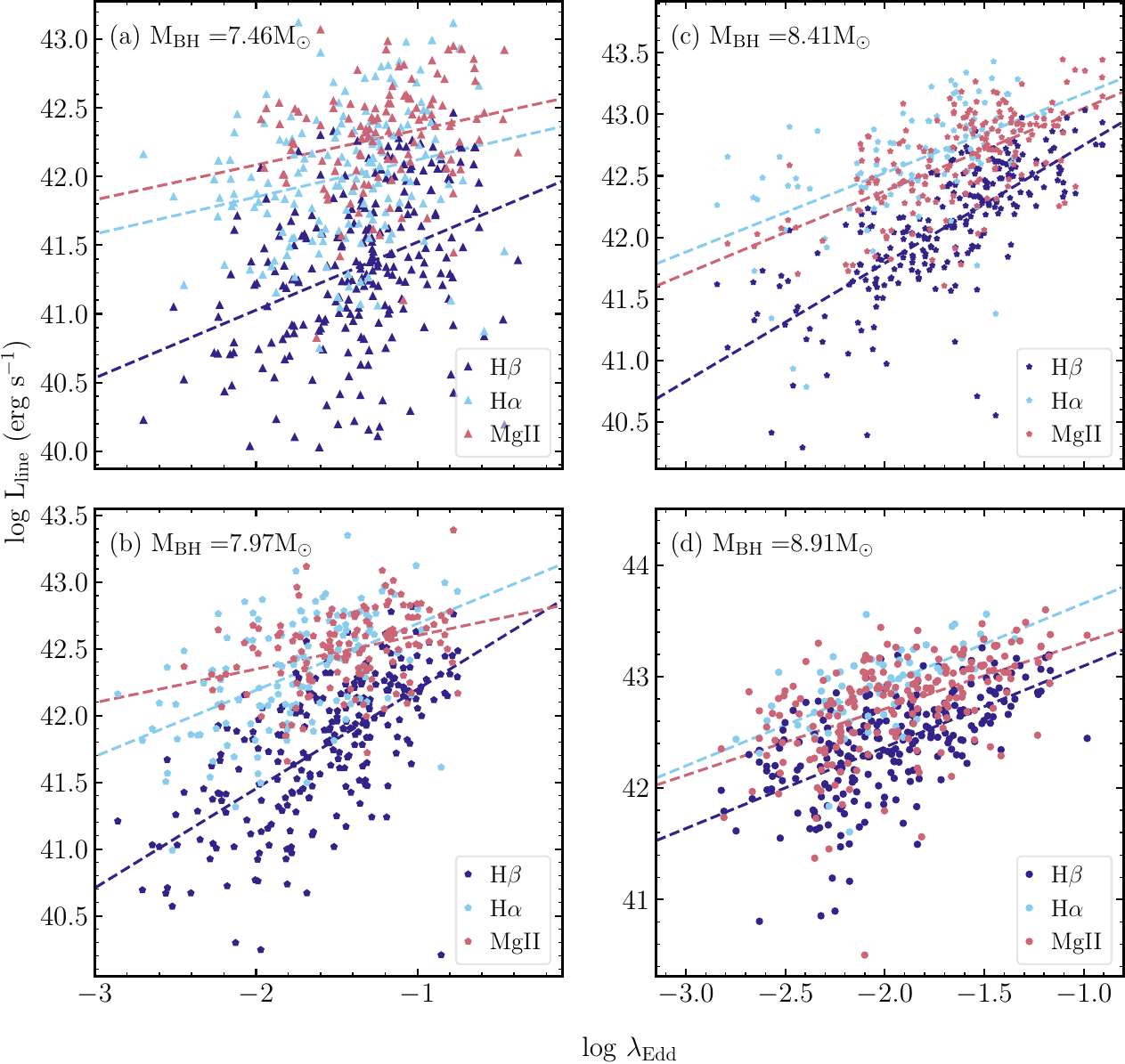}
\caption{
   Correlation between broad emission-line luminosities ($L_{\rm line}$) and the Eddington ratio ($\lambda_{\rm Edd}$) for all CL-AGNs, binned by black hole mass ($M_{\rm BH}$). Each panel corresponds to a different $M_{\rm BH}$ bin, as labeled. The emission lines H$\beta$ (dark blue), H$\alpha$ (light blue), and Mg\,\textsc{ii} (red) are shown individually, with linear regression fits overplotted as dashed lines. Error bars are omitted here for clarity due to the large number of points.}
\label{fig:line_eddtion_CLAGN}
\end{figure*}

\subsection{broad line Luminosity and Eddington Ratio Correlation}

CL-AGNs offer direct time-domain evidence linking broad line luminosity to accretion rate. During state transitions, their Eddington ratios vary significantly, resulting in the emergence or disappearance of broad lines. While normal AGNs exhibit a static correlation between $L_{\rm line}$ and $\lambda_{\rm Edd}$, CL-AGNs trace the full dynamical range, showing order-of-magnitude variability in both quantities.

Figure~\ref{fig:line_eddtion} shows the correlation between broad emission-line luminosities ($L_{\rm line}$) and the Eddington ratio ($\lambda_{\rm Edd}$) in normal Type~1 AGNs, grouped by black hole mass ($M_{\rm BH}$). A clear positive trend is observed across all mass bins, consistent with expectations from photoionization models. Figure~\ref{fig:line_eddtion_CLAGN} presents the same correlation for CL-AGNs, also split by $M_{\rm BH}$. Although a positive trend remains, the CL-AGN sample exhibits significantly greater scatter, reflecting their transient nature and extreme variability.

Both populations show a strong link between broad line luminosity and $\lambda_{\rm Edd}$: higher accretion rates generate a more luminous ionizing continuum, leading to stronger broad emission. RM confirms that broad line flux tracks continuum changes. While the relation is not perfectly linear, normal Seyferts and quasars with higher $\lambda_{\rm Edd}$ tend to have brighter H$\alpha$, H$\beta$, and Mg\,\textsc{ii} lines than lower-$\lambda_{\rm Edd}$ counterparts at similar mass, reflecting increased ionizing output.

These results support the idea that the BLR is sustained only above a critical accretion threshold \citep{Elitzur2014}. In CL-AGNs, broad line and continuum luminosities vary in tandem: as the continuum dims, broad lines fade by comparable factors, sometimes vanishing entirely in “turn-off” events, and re-emerging during “turn-on” phases. Notably, even Mg\,\textsc{ii}—which is typically less responsive in steady-state AGNs—shows strong variability in CL-AGNs when continuum changes are large.

Transitions between Type~1 and Type~2 spectral states in CL-AGNs are consistently associated with substantial shifts in $\lambda_{\rm Edd}$ \citep{Yang2015, Noda2018, Jana2025}. In their dim state, CL-AGNs typically exhibit very low accretion rates, with a median $\lambda_{\rm Edd} \approx 0.01$ \citep{Panda2024, Guo2025}. This $\sim$1\% level appears to mark a critical threshold below which broad line emission often disappears, though the exact value depends on black hole mass.

\subsection{Mass-dependent Trends in broad line Evolution}

While accretion rate primarily governs the presence of broad lines, black hole mass introduces important secondary effects. Across all $M_{\rm BH}$ bins, we observe a consistent trend: higher $\lambda_{\rm Edd}$ corresponds to strong broad line emission, while low $\lambda_{\rm Edd}$ ($\lesssim 10^{-2}$) generally results in weak or undetectable broad lines (Figures~\ref{fig:normalAGN_stacks}, \ref{fig:line_eddtion}, and \ref{fig:line_eddtion_CLAGN}).

Interestingly, the threshold at which broad lines disappear varies with mass. In high-mass systems ($\log M_{\rm BH}/M_\odot \gtrsim 8.4$), H$\alpha$ and Mg\,\textsc{ii} remain detectable even at $\log \lambda_{\rm Edd} \lesssim -2$. In contrast, lower-mass AGNs ($\log M_{\rm BH}/M_\odot \sim 7.1$) begin to lose broad lines—especially H$\beta$—near $\log \lambda_{\rm Edd} \sim -1$. This systematic shift suggests that the critical $\lambda_{\rm Edd}$ for sustaining the BLR is mass-dependent.

Although the $\sim$1\% threshold remains a useful benchmark, its manifestation varies with mass:

\begin{itemize}
    \item \textbf{High-mass black holes}: Systems with $M_{\mathrm{BH}} \gtrsim 10^8$–$10^9\,M_\odot$ can maintain significant luminosity even at modest $\lambda_{\rm Edd}$. However, once $\lambda_{\rm Edd}$ falls below $\sim10^{-2}$–$10^{-3}$, broad lines often disappear, leading to true Type~2 AGNs \citep{Green2022,Zeltyn2024,Guo2025}. This helps explain why CL events are rare among luminous quasars: a larger drop in accretion is required to cross the threshold.

    \item \textbf{Low-mass black holes}: AGNs with $M_{\mathrm{BH}} \sim 10^7$–$10^{7.5}\,M_\odot$, including many NLSy1s, often operate at high $\lambda_{\rm Edd}$ and exhibit strong broad lines and Fe\,\textsc{ii} emission. However, even modest reductions in accretion can push these systems below the BLR-sustaining threshold (around $\log \lambda_{\rm Edd} \sim -1.25$), leading to CL transitions \citep{Panda2024}.
\end{itemize}

\section{Discussion}
\label{sec:discussion}

\subsection{BLR Stratification Structure}

The observed differential fading sequence of H$\beta$, H$\alpha$, and Mg\,\textsc{ii} can be naturally interpreted within the framework of the BLR structure and standard photoionization physics. According to the classical LOC model, the BLR comprises gas clouds spanning a broad range of radii and densities, with each emission line predominantly originating from regions where physical conditions favor its formation \citep{Baskin2018}. As the ionizing continuum luminosity declines, the innermost BLR—dominated by high-ionization gas illuminated by energetic photons—undergoes the most significant reduction in emission. H$\beta$, which requires higher excitation energy, arises from the innermost regions and therefore fades first. In contrast, H$\alpha$, with a lower ionization potential, originates from cooler or more extended regions and diminishes more gradually. Mg\,\textsc{ii}, a low-ionization line with an ionization potential of $\sim$15\,eV, is expected to form at even larger BLR radii, typically beyond the H$\beta$ zone.

This radial stratification naturally gives rise to the observed fading sequence: H$\beta$ $\rightarrow$ Mg\,\textsc{ii} $\rightarrow$ H$\alpha$. Notably, \citet{guoh2020} used {\tt Cloudy} photoionization simulations within the LOC framework to demonstrate that Mg\,\textsc{ii} emission remains prominent even as the ionizing continuum dims, due to its production in the outer BLR that is less affected by moderate drops in photon flux. These results provide a physical basis for the empirical observation that some CL quasars retain broad Mg\,\textsc{ii} emission even after the disappearance of broad Balmer lines—a pattern also evident in our CL-AGN sample.

This interpretation is further supported by RM studies, which independently reveal the radial stratification of the BLR: $R_{\rm H\beta} \lesssim R_{\rm MgII} \lesssim R_{\rm H\alpha}$ \citep{Stone2024,Shen2024}. In such campaigns, H$\beta$ exhibits the shortest time delays, indicative of an inner BLR origin, while Mg\,\textsc{ii} and H$\alpha$ respond on longer timescales, implying larger emission radii. Together, these observational and theoretical results reinforce a stratified BLR structure, where each line's variability reflects its ionization requirements and spatial location.

Two physical mechanisms contribute to this behavior. First, H$\alpha$ emissivity remains relatively efficient even when the ionization parameter drops, as it originates from transitions between low-lying energy levels ($n=3 \rightarrow 2$) and can be generated in partially ionized, cooler gas \citep{netzer2015}. In contrast, H$\beta$ ($4861\,\text{\AA}$, $n=4 \rightarrow 2$) requires more energetic excitation and is typically about five times weaker than H$\alpha$ under standard conditions. As the total line luminosity decreases, H$\beta$’s equivalent width diminishes more rapidly and may become undetectable against the continuum long before H$\alpha$.

Second, dust extinction and optical depth effects within the BLR preferentially suppress shorter-wavelength emission. As the BLR becomes increasingly dust-rich or optically thick—especially in the context of a dissipating disk wind—blue and UV photons are attenuated more strongly, leading to a preferential loss of H$\beta$ and Mg\,\textsc{ii} relative to the redder H$\alpha$. The final stages of BLR evolution are thus often marked by residual broad H$\alpha$ emission with narrower velocity width, consistent with type 1.9 AGN spectra. Continued accretion decline eventually suppresses even H$\alpha$, resulting in a complete disappearance of broad lines and a transition to a type 2 AGN. This sequence—H$\beta$ $\rightarrow$ H$\alpha$/Mg\,\textsc{ii} $\rightarrow$ none—is fully consistent with disk-wind models \citep{Elitzur2014}, in which both the size and optical depth of the BLR contract, leaving only the most resilient low-ionization gas before total quenching.

\subsection{Outliers Diverge from the Evolutionary Sequence}
\label{sec:outlier}
As shown in Figures~\ref{fig:line_luminosity_CLAGN}, we identify a subset of CL-AGNs where broad Mg\,\textsc{ii} emission persists even after the disappearance of H$\alpha$. This deviation from the expected fading sequence—H$\beta \rightarrow$  Mg\,\textsc{ii} $\rightarrow$ H$\alpha $ —challenges simple photoionization models based on a radially stratified BLR. Within the standard LOC framework, Mg\,\textsc{ii} is expected to originate in the outer BLR and therefore fade later than the Balmer lines as the ionizing continuum declines. However, the presence of strong Mg\,\textsc{ii} emission in the complete absence of H$\alpha$ indicates that additional physical mechanisms must be considered. If H$\beta$ and H$\alpha$ arise from regions separated by several to tens of light-days, then for changing-look transitions on year-long timescales, the probability of observing H$\alpha$ after H$\beta$ has vanished is only a few percent. Thus, light-travel delays alone cannot account for the observed phenomenon.

Similar cases have been reported previously. \citet{guohengxiao2019} and \citet{Roig2014} identified AGNs with strong Mg\,\textsc{ii} but weak or absent Balmer lines, termed ``Mg\,\textsc{ii} emitters.” Such observations underscore the limitations of simplified photoionization models and point to more complex BLR physics.

A plausible explanation involves differences in excitation mechanisms. While Balmer lines are primarily powered by radiative recombination, Mg\,\textsc{ii} is a resonance line strongly influenced by collisional excitation \citep[e.g.,][]{Guo2020, Sniegowska2020}. At high BLR densities, Mg\,\textsc{ii} can remain strong even when the ionizing continuum falls below the level required for Balmer recombination. Among the Balmer lines, H$\alpha$ is also more easily excited by collisions than H$\beta$, making its persistence more likely under conditions where collisional processes become important. 

This framework suggests that fading sequences are governed not only by radial stratification but also by excitation physics and thermal response. When the ionizing source turns off, H$\beta$—most sensitive to photoionization—disappears first. H$\alpha$, benefiting from collisional excitation, persists longer, and Mg\,\textsc{ii}, dominated by collisional processes, can remain detectable until the BLR gas cools, on timescales comparable to the thermal cooling time rather than the light-crossing time. 

This interpretation yields testable predictions: in turn-off CL-AGNs, the expected sequence is H$\beta \rightarrow$ H$\alpha \rightarrow$ Mg\,\textsc{ii}, while in turn-on events, Balmer lines should appear promptly, with Mg\,\textsc{ii} emerging later as collisional excitation strengthens with rising gas temperature. Thus, in brightening states, H$\alpha$ and H$\beta$ precede Mg\,\textsc{ii}, whereas in fading states, Mg\,\textsc{ii} lingers last. 

Alternative scenarios invoke BLR geometry and SED evolution. The BLR may be clumpy and anisotropic rather than homogeneous, leading to selective obscuration of Balmer-emitting regions, or the ionizing SED may harden or soften in ways that favor Mg\,\textsc{ii} production over Balmer excitation. Optical depth and radiative transfer effects may also contribute. 

Additional evidence for such complexity comes from reverberation mapping (RM) studies. Some RM campaigns report $\tau_{\rm Mg\,II} > \tau_{\rm H\alpha}$, consistent with Mg\,\textsc{ii} arising at larger radii \citep{Li2017}, while others find similar or even inverted size ordering \citep[e.g.,][]{Bentz2013, Homayouni2020, Yu2021, Cho2023, Li2024}. Long-term monitoring, such as the five-year study of PG~2130+099 by \citet{yao2024}, reveals significant temporal evolution in line lags, suggesting that BLR structure is dynamic and responds to changes in accretion rate and in the of the central engine—i.e., the UV–EUV/soft X-ray continuum from the disk+corona that illuminates the BLR and modulates line responsivities and the emissivity-weighted radius \citep{Ferland2020,Panda2022Fr}.

Overall, these findings highlight the need for time-dependent BLR models that incorporate photoionization, collisional excitation, and thermal physics within an evolving, anisotropic geometry. Future time-resolved spectroscopy of turn-on and turn-off CL-AGNs will be critical for testing these predictions and refining our understanding of BLR physics.

\subsection{Low-Mass Systems: Deviations and Challenges}
\label{sec:lowmass}

Figure~\ref{fig:line_luminosity2} and Figure~\ref{fig:line_luminosity} reveals a significant deviation in the low-luminosity regime of the broad line versus continuum relation ($L_{\rm line}$–$L_{5100}$). At the lowest continuum luminosities, the broad line luminosity falls below the extrapolated trend established by higher-luminosity sources—i.e., low-$L_{5100}$ AGNs exhibit systematically weaker broad-emission lines than expected. A similar trend is observed in the Eddington ratio plane (Figure~\ref{fig:line_eddtion}, panel~a), where the lowest $M_{\rm BH}$ bins display a flattening of the $L_{\rm line}$–$\lambda_{\rm Edd}$ relation at high $\lambda_{\rm Edd}$, suggesting saturation of line emission at extreme accretion rates. Together, these patterns indicate a breakdown of standard photoionization scaling relations in low-mass black hole systems under certain physical conditions.

Several factors may contribute to these deviations:  
(i) \textbf{BLR covering factor.} The fraction of ionizing photons intercepted by the BLR may be reduced in low-luminosity or low-accretion states. A diminished covering factor would naturally suppress $L_{\rm line}$, particularly in CL-AGN ``turn-off'' phases where the BLR partially disappears.  
(ii) \textbf{Accretion disk structure.} At high $\lambda_{\rm Edd}$, the accretion flow may transition to a slim disk geometry—optically thick and geometrically puffed-up—which alters the emergent ionizing spectrum. Such disks radiate inefficiently at high $\dot{M}$, with the ionizing luminosity increasing only weakly due to radiation trapping and advection. Moreover, the inflated inner disk can self-shadow the BLR, shielding clouds from ionizing photons and leading to a plateau in $L_{\rm line}$ despite increasing $\lambda_{\rm Edd}$ \citep{Wang2013}.  
(iii) \textbf{Geometrical inclination.} Viewing angle effects may influence both the observed continuum and line strengths. Low-mass AGNs, particularly narrow-line Seyfert 1 galaxies (NLSy1s), are thought to be viewed preferentially face-on. While this minimizes Doppler broadening, it may also affect apparent luminosities due to orientation-dependent obscuration or anisotropic disk emission \citep{Panda2019,Panda2021}. Such effects could introduce scatter or bias in the $L_{\rm line}$–$L_{5100}$ relation.

Importantly, the low-mass subsample is dominated by NLSy1 galaxies, which are known to deviate systematically from the broader quasar population. These sources typically host small black holes ($M_{\rm BH} \sim 10^6$–$10^7\,M_\odot$) but radiate at near- or super-Eddington rates. As a result, they occupy a distinct locus in the $M_{\rm BH}$–$L_{\rm bol}$ plane and often display elevated $\lambda_{\rm Edd}$. \citet{Frederick2019} showed that NLSy1s lie above the main sequence of broad line AGNs, exhibiting higher luminosities at a given $M_{\rm BH}$. This behavior reflects their extreme accretion physics: NLSy1s frequently display strong Fe\,\textsc{ii} emission and steep soft X-ray spectra, consistent with high $\lambda_{\rm Edd}$, and often drive powerful winds \citep{Marziani2018, Panda2024Fr}. \citet{Mullaney2013} further linked high-$\lambda_{\rm Edd}$ AGNs to enhanced outflow signatures, implying altered BLR and NLR conditions in this regime. Thus, the prevalence of NLSy1s in the low-mass sample likely accounts for the observed departures from canonical $L_{\rm line}$–$L_{5100}$ and $L_{\rm line}$–$\lambda_{\rm Edd}$ relations.

Finally, the saturation of $L_{\rm line}$ at high Eddington ratios mirrors results from RM studies of super-Eddington AGNs. These studies have found that high-$\lambda_{\rm Edd}$ sources exhibit shorter-than-expected H$\beta$ lags for their luminosity, indicating that the BLR size increases more slowly with $L$ than in sub-Eddington systems. \citet{Du2016} proposed a revised H$\beta$ radius–luminosity relation for high-accretion-rate AGNs, showing that beyond a critical $\dot{M}$, the BLR size saturates due to geometric and radiative constraints—namely, the slim disk's vertical expansion and radiation trapping. In this scenario, further increases in accretion rate yield diminishing returns in ionizing luminosity and, consequently, in recombination-line output. Our observed flattening of the $L_{\rm line}$–$\lambda_{\rm Edd}$ relation among low-mass AGNs is consistent with this picture. In NLSy1-like systems, the BLR may be physically and geometrically constrained in its ability to respond to continuum variations. Once fully ionized or structurally limited, the BLR produces only marginal gains in line luminosity despite rising accretion power. These results underscore the inadequacy of simple photoionization models and point to the necessity of incorporating additional parameters—such as BLR geometry, cloud column density, and anisotropic illumination—when modeling line emission in the low-mass, high-$\lambda_{\rm Edd}$ regime \citep{Panda2021}.

\section{Conclusion}
\label{sec:conclusion}

We present the first statistical evidence for a stratified evolutionary sequence in the BLR of CL-AGNs, based on a redshift-selected subsample ($0.35 < z < 0.45$) drawn from the 561 CL-AGNs identified in \citet{Guo2025}. This subset enables simultaneous coverage of the H$\alpha$, H$\beta$, and Mg\,\textsc{ii} emission lines in DESI spectra. By comparing these CL-AGNs to a control sample of normal Type~1 AGNs at similar redshifts, we examine the dependence of broad line luminosities on the Eddington ratio and uncover a well-defined fading sequence: as continuum luminosity or $\lambda_{\mathrm{Edd}}$ decreases, broad H$\beta$ fades first, followed by Mg\,\textsc{ii}, and finally H$\alpha$.

This empirical sequence is consistent with predictions from photoionization models and supports a physically stratified BLR, in which line responsivity is governed by both ionization potential and radial location. Our results further reveal that broad line visibility is tightly linked to accretion rate, with a critical threshold near $\lambda_{\mathrm{Edd}} \sim 0.01$ below which the BLR structure collapses. Additionally, we find that this threshold is not universal, but may depend on black hole mass: more massive black holes are capable of sustaining broad line emission at lower accretion rates.

The strong agreement between our findings and those from RM studies suggests that the BLR evolutionary sequence we identify—H$\beta$ $\rightarrow$ Mg\,\textsc{ii} $\rightarrow$ H$\alpha$—is a general and robust phenomenon across AGNs. It reinforces a unified physical picture: as the accretion power of an AGN diminishes, the BLR contracts, and weak ionization lines disappear first. This pattern, reproducible across samples and methodologies, lends strong support to accretion-driven models of CL-AGN variability.

Our analysis advances prior studies by systematically quantifying the multi-line evolution within individual CL-AGNs, thereby providing a more physically grounded framework that links BLR structure to the state of the accretion flow. This not only clarifies previously reported CL-AGN behaviors but also offers predictive power. For example, an AGN observed in a dim state exhibiting only broad Mg\,\textsc{ii} emission may have recently lost its broad H$\beta$ and H$\alpha$ due to a drop in $\lambda_{\rm Edd}$—and could potentially recover them should the accretion rate increase. Such insights are essential for designing time-domain monitoring campaigns and refining theoretical models of AGN variability, including those involving disk instabilities or transitions to radiatively inefficient accretion modes.

\section*{acknowledgements}

We thank the anonymous referee for the constructive comments and suggestions, which helped us to improve the clarity and quality of this work. Y.R.L. acknowledges financial support from the the National Natural Science Foundation of China (12273041), from the National Key R\&D Program of China (2023YFA1607904), and from the Youth Innovation Promotion Association CAS.  W.J.G. acknowledges financial support from  National Natural Science Foundation of China (NSFC) under grant No. 12503019.  VAF acknowledges funding from a United Kingdom Research and Innovation grant (code: MR/V022830/1). M.S. acknowledges support by the State Research Agency of the Spanish Ministry of Science and Innovation under the grants 'Galaxy Evolution with Artificial Intelligence' (PGC2018-100852-A-I00) and 'BASALT' (PID2021-126838NB-I00) and the Polish National Agency for Academic Exchange (Bekker grant BPN/BEK/2021/1/00298/DEC/1). This work was partially supported by the European Union's Horizon 2020 Research and Innovation program under the Maria Sklodowska-Curie grant agreement (No. 754510). SP is supported by the international Gemini Observatory, a program of NSF NOIRLab, which is managed by the Association of Universities for Research in Astronomy (AURA) under a cooperative agreement with the U.S. National Science Foundation, on behalf of the Gemini partnership of Argentina, Brazil, Canada, Chile, the Republic of Korea, and the United States of America.  J.M.W. acknowledges financial support from the National Key R\&D Program of China (2021YFA1600404) and from the National Natural Science Foundation of China (12333003). 

This material is based upon work supported by the U.S. Department of Energy (DOE), Office of Science, Office of High-Energy Physics, under Contract No. DE–AC02–05CH11231, and by the National Energy Research Scientific Computing Center, a DOE Office of Science User Facility under the same contract. Additional support for DESI was provided by the U.S. National Science Foundation (NSF), Division of Astronomical Sciences under Contract No. AST-0950945 to the NSF’s National Optical-Infrared Astronomy Research Laboratory; the Science and Technologies Facilities Council of the United Kingdom; the Gordon and Betty Moore Foundation; the Heising-Simons Foundation; the French Alternative Energies and Atomic Energy Commission (CEA); the National Council of Science and Technology of Mexico (CONACYT); the Ministry of Science and Innovation of Spain (MICINN), and by the DESI Member Institutions: \url{https://www.desi.lbl.gov/collaborating-institutions}.

Any opinions, findings, and conclusions or recommendations expressed in this material are those of the author(s) and do not necessarily reflect the views of the U. S. National Science Foundation, the U. S. Department of Energy, or any of the listed funding agencies.

The authors are honored to be permitted to conduct scientific research on Iolkam Du’ag (Kitt Peak), a mountain with particular significance to the Tohono O’odham Nation.

\bibliographystyle{aasjournal}
\bibliography{main}
\end{document}